\documentclass[11pt]{article}
\usepackage{amsmath}
\usepackage{epsfig,bm}
\def\b{\begin}
\def\e{\end}
\def\t{thebibliography}
\def\bi{\bibitem}
\def\be{\b{equation}}
\def\ee{\e{equation}}

\def\eq{\equiv}

\begin{document}

\title{INFORMATION CONSERVATION, ENTROPY INCREASE
AND THE STATISTICAL IRREVERSIBILITY FOR AN ISOLATED SYSTEM }

\author{Qi-Ren Zhang
\\Department of Technical Physics, Peking University , Beijing,100871,
China}

\maketitle

\vskip0.3cm

\begin{abstract}
We consider the statistical irreversibility and its compatibility
with the reversible dynamics. The role played by the observation is
analyzed in detail. It makes our previous proof for the second law
of thermodynamics clearer.  On this basis, we emphasize the
importance and wide applicability for the second law of
thermodynamics. A new form of physics with this law being
substituted by the principle of information conservation is
suggested. By the way, we also solve the paradox of Schr\"odinger
cat, and show that the universe will not go to the so-called heat
death spontaneously.

\vskip 0.5cm
\noindent PACS: 05.30.Ch, 05.70.-a, 03.65.-w

\noindent Keywords: Information conservation, Information change in
observation, Entropy increase, Statistical irreversibility

\end{abstract}

\bigskip

\section{Introduction}
We proved the second law of thermodynamics exactly\cite{1,2}, from
the principle of information conservation, which is exactly valid
both in classical and quantum dynamics. However the proof was purely
mathematical. Here we would complete it with a physical analysis to
make the proof qualitatively understandable. Then we may say that
the problem of understanding this law on the basis of dynamics and
statistics is completely solved. The analysis shows, if the dynamics
is reversible, the irreversibility of a macroscopic process may
arise because of the change of information and the corresponding
change of statistic ensemble in the observation. We call this kind
of irreversibility statistical.

\section{Information and its change in observations}
For any probability distribution $[W_n]$ one may define its
information \b{eqnarray}{\cal I} =\sum_n W_n \ln W_n\;
.\label{1}\e{eqnarray} After an observation, the probability
distribution and therefore the information changes. For an example,
in a complete measurement one finds that the possibility $n=n_0$ is
realized, and all other possibilities of $n\ne n_0$ disappear. The
probability distribution becomes $W_{n_0}=1$, $W_{n\ne n_0}=0$, and
the information becomes ${\cal I}_0 =0$.  The observer gains the
information ${\cal I}_0 -{\cal I} = -{\cal I} =-\sum_n W_n \ln W_n
\geq 0 $. There are also observations, in which the observer loses
information.

Consider a set $[|n\rangle]$ of independent states  for a system, in
which $|n\rangle$ is the $n$-th state in the set. The set, with a
probability distribution $[W_n]$ defined on it, is called a
statistic ensemble of the system. Substituting the distribution
$[W_n]$  into (\ref{1}), the obtained $\cal I$ is called the
information of the system. Divide the system into subsystems. The
probability distributions on the sets of independent states for
subsystems are obtained by reducing the probability distribution of
the system, and the information of subsystems is defined. There is
an exact mathematical inequality\cite{1,3} \be \sum_i {\cal I}_i
\leq {\cal I} \;  , \label{2}\ee in which ${\cal I}_i$ is the
information of the $i$-th subsystem, the summation is over all
subsystems, and the equality holds if and only if the probability
distribution for the system is factorized into a direct product of
the probability distributions for subsystems. When one observes the
system by measuring its subsystems separately, he ignores the
correlation between the subsystems, the probability distribution for
the system is therefore factorized. He gets the information ${\cal
I}_0 =\sum_i {\cal I}_i$ for the system. According to (\ref{2}), the
observer usually loses information in this kind of observation.

\section{Information conservation and the entropy increase}
The states of the system evolve in time according to the dynamics.
For an isolated system, independent states always evolve
continuously into independent states. This is true both in quantum
and in classical dynamics. It means the probability distribution on
independent states does not change in time, only the states on which
the probability distribution is to be defined may be changed.
Therefore the information of the isolated system does not change in
time either. This is the information conservation. It is a
fundamental principle both for quantum and classical physics.

Entropy is an extensive variable. The entropy of a system is
therefore defined by the sum of entropies of its individual
subsystems. The entropy of an individual subsystem is defined by its
negative information in the unit of Boltzmann constant. The entropy
of the system in this unit is therefore defined by\cite{l} \be S \eq
-\sum_i {\cal I}_i\;  ,\label{3}\ee in which ${\cal I}_i$ is the
information of the $i$-th subsystem, and the summation is over all
of its subsystems. From (\ref{2}) we see that the entropy and the
information for a system are related by $S \ge -{\cal I}$.  If one
wants to determine the entropy of the system at an initial time
$t_0$, he has to measure the entropy of each individual subsystem.
In this observation he loses the correlation information between the
subsystems. The obtained entropy at this time is therefore $S(t_0) =
-\sum_i {\cal I}_i (t_0) = -{\cal I}(t_0)$. From time $t_0$ to time
$t$, the system evolves according to its dynamics with information
conservation. It means that the information of the system at time
$t$ is ${\cal I} (t) = {\cal I}(t_0)$, and the entropy of the system
at time t is therefore \be S (t) \ge -{\cal I} (t) = -{\cal I} (t_0)
= S (t_0)\; . \ee  It shows that the entropy of an isolated system
strictly increases between two successive observations, unless the
probability distribution for the system keeps being factorized at
time t, it is that there is no correlation between these subsystems.
In the later case the entropy does not change, and the system is in
equilibrium. This is the principle of entropy increase. The
factorization condition of probability distribution is just the
start point for the derivation of Gibbs distribution for equilibrium
systems\cite{l}.

\section{Ensemble change and the statistical irreversibility}
The observer gains or loses information in an observation, therefore
he has to redefine the statistic ensemble after the observation,
just like a general changes his estimation on the war situation by
receiving a telegram from the front. In our case, the ensemble of
the system is considered. The states in ensemble A at time $t_0$
evolve into the states in ensemble B at time $t$, according to the
dynamics of the system. If the dynamics is reversible, the time
inverse ensemble B' of B, obtained by changing every state in B to
its time inverse state, evolves from time $-t$ to time $-t_0$ into
the time inverse ensemble A' of A. However, an observation on the
system at time $t$ changes the ensemble from B to C, its time
inverse ensemble C' at $-t$ does not evolve into the ensemble A' at
$-t_0$ in general. This is the statistical irreversibility. We see
that it is perfectly compatible with the dynamical reversibility.

The consideration of observation in quantum physics needs more
caution, because of the so-called wave packet collapse. It is
assumed that a superposition of partial waves suddenly collapses
into one of its partial waves when a measurement is performed, and
the process is irreversible. Since there is not a clear partition
between the measured system, the measuring instrument, and the
observer, one may assume that these three parts consist a single
macroscopic system. When one considers the wave packet collapse of
this kind of system, he immediately falls into the discomfiture of
the so called Schr\"odinger cat. The way to escape from this
discomfiture is to recognize that the wave function is a
probabilistic description of the system. Every partial wave
corresponds to a set of possible properties, and the relative
strengths of the partial waves denote the probability distribution
on the sets of possible properties. When these partial waves
superpose in a common configuration space, they interfere each
other. But it becomes more and more difficult to make two objects
superpose when the dimension of space increases. To understand it,
let us consider a model, in which a point moves in a $f$-dimensional
space of volume $L^f$, and ask the probability of the point passing
through a small box of volume $l^f$ in this space, when it moves
from one side of the space to another side. This model may be used
to simulate the coincidence of two partial waves in a
$f$-dimensional configuration space. If the motion is in random, the
coincidence probability is $(l/L)^{f-1}$, which decreases
exponentially with the dimension $f$; and the time one has to wait
for seeing this coincidence is as long as $T\ge (L/l)^{f-1}L/c$,
which increases exponentially with $f$, $c$ is the speed of light.
Take $L=1$micron, it is not big, and $l/L=0.01$, it is not too
small. We see, for $f =18$, which is the degrees of freedom for 6
particles in a 3-dimensional space, the coincidence probability is
$10^{-34}$, it is practically zero; and $T\ge 3\times 10^{19}$sec,
it is longer than the universe age and means people have never seen
this kind of coincidence since the birth of the universe, or it has
never happened. Based on this understanding, one may define the
measurement on a system to be an interaction between this system and
a macroscopic system, usually (but not always) composed of the
measuring instrument and the observer, which makes the wave packet
of the total system decompose into a number of partial waves
separated by macroscopic distance in their configuration space of
numerous degrees of freedom. This is a purely objective process
governed by the usual physical laws. Since these partial waves are
practically impossible to meet each other once again, they cannot
interfere each other any more. At the end, the observer finds that
the properties corresponding to one of those partial waves is
realized, and this result will appear repeatedly in his following
observations on the same system. The assumption of a physical wave
packet collapse is therefore unnecessary. As in the classical case,
The observer has to change his information and the statistic
ensemble for the system, from that described by the original wave
packet to that described by the partial wave, which he read off from
the measuring instrument. This change leads to the statistical
irreversibility. By the way, the paradox of Schr\"odinger cat is
solved. Before the observer sees it, the cat was either alive or
dead after the corresponding partial waves being macroscopically
separated and practically not able to interfere each other any more.
Nothing is strange. This argument offers a concrete explanation on
the role played by observation in above proof for the principle of
entropy increase at the quantum level.

\section{Meanings and applications}
The sum ${\cal I}_0 \eq \sum_i {\cal I}_i$ of the information of all
individual subsystems is thought to be a measure of  order for the
system. Correspondingly, equation (\ref{3}) shows that the entropy
is a measure of disorder for the system. If we regard the whole
universe as an isolated system, the second law of thermodynamics
seems to show that the universe shall become more and more
disorderly. This is the so-called heat death of the universe, and is
worried by many physicist. In the following, we will see this is a
paradox only, and may be resolved. As shown in our proof of the
second law of thermodynamics, when one observes the system and finds
the entropy increase, he has redefined the statistic ensemble under
research, and changed the information of the system from ${\cal I}$
to ${\cal I}_0$ . The decrease of ${\cal I}_0$ compared with its
value found in the last observation makes the entropy increase.
However, the redefinition of the ensemble by the observer is
subjective. For the objective evolution of the original ensemble,
the information $\cal I$ is conserving. The entropy increase, it is
the decrease of ${\cal I}_0$ , is compensated by the increase of the
correlation information \be {\cal I}_c \eq {\cal I} - {\cal I}_0
\label{7}\ee between individual subsystems of the system, although
it is ignored in the presentation of the second law. For a given
ensemble in its time evolution, the more suitable measure of the
order should be the (total) information $\cal I$ . It does not
change in time. From this view we see, that an isolated system will
not become more and more disorderly. The universe, even if it may be
thought to be an isolated system, will not approach the absolute
disorder. The paradox of heat death is therefore resolved.

The second law of thermodynamics may be derived mathematically from
the information conservation, means it is absolutely true in the
present physics. There are a lot of conservation laws in the present
physics, examples are energy conservation, momentum conservation,
angular momentum conservation, ...and so on, they are of firm
foundation and related with various symmetries in time, space and
matter. Information conservation is also related with the symmetry
of the physics. It is the symmetry under the unitary transformation
in quantum theory, or the symmetry under the canonical
transformation in classical theory. However, these two symmetries
are the most fundamental symmetries in the quantum and classical
theories respectively. It means that one may image a fundamental
quantum/classical theory without time translation symmetry and
therefore is not energy conserving, but any fundamental
quantum/classical theory must be symmetric under the
unitary/canonical transformation and therefore is information
conserving. In this sense, the principle of information
conservation, and therefore the second law of thermodynamics, is the
most fundamental law in physics. The foundation of the information
conservation is even wider than the unitary/canonical symmetry. As
we have shown before, if independent states of a system evolve
continuously into independent states, the dynamics is information
conserving. One may wonder what will be the future (post quantum and
post relativistic) physics, which kind of law will still hold in it?
From the above analysis we see, the most hopeful one is the
information conservation, or the second law of thermodynamics.

The above argument also offers a new form of representation for
physics, in which, among the fundamental principles, the second law
of thermodynamics is substituted by the principle of information
conservation. The second law itself becomes a deduced inequality. At
least, from the view point of aesthetics, this form is attractive.
It is more symmetric. All quantitative fundamental principles are
presented in the form of equations. All inequalities are deduced.
Moreover, this form may be easier to use in the research of ensemble
evolution between two successive observations, especially on the
transfer of information from ${\cal I}_0$ to ${\cal I}_c$ . I am
also happy to note, that this form has already been dreamed by some
one else\cite{4}.

Our proof of entropy increase relies only on the information
conservation, therefore is valid not only for physics, but also for
any statistical process, in which the information conservation is
true.  In this kind of process, the entropy of an isolated system
always increases until reaches the equilibrium. This  is important
since the entropy increase may point out the direction of
development. To find out these kinds of processes in biology,
engineering, social science, and financial science would therefore
be interesting

\section{Discussions}
Some people may be surprised, since they feel the second law of
thermodynamics should be approximate only. They could not believe an
exact irreversible law may be founded on a reversible dynamics.
There are indeed papers discuss its possible violation. One example
is connected with the fluctuation theory\cite{5,6}. This is an
interesting and valuable theory, but I am afraid that authors'
declaration on the violation of the second law of thermodynamics may
be wrong. The appearance of the entropy consuming trajectories is
declared to be the evidence of the second law violation. However,
the second law requires that the entropy does not decrease only for
isolated systems. But systems which move along the entropy consuming
trajectories in their examples are not isolated, but interacting
with the surrounding medium.

The only possible objection against the above argument may be that,
the observation contains subjective element, and the content of
physics should be purely objective. This is a classical point of
view in philosophy. We would say yes, physicists make every effort
try to find out the truth of the objective world, but the only way
to do this is by their observation. To extract the objective truth
from observation, one needs a theory of the observation, at least he
has to know the role played by the observation in his research. One
element in observation is the physical interaction between the
observed system, the measuring instrument, and the observer. This is
purely objective, and is governed by physical laws. Another element
is the decision of the observer on the kind of observation to be
performed, and the changing of his mind after the observation. This
is subjective. The later includes the possible redefinition of the
statistical ensemble for further research. These two elements are
nature and necessary for an observation. The above argument shows
the role played by the observation in understanding the statistical
irreversibility. Behind it one sees the purely objective reversible
dynamics.

Some people show the violation of the second law by use of effective
theories, such as Fokker-Planck equation, and by use of various
approximate methods. This is not reliable. Theoretically, to prove
or disprove a law as fundamental as the second law of
thermodynamics, one has to use the fundamental theory and rigorous
mathematical method.

The early attempt of proving the second law of thermodynamics was
made by Boltzmann in nineteenth century. In the course, he proposed
and proved his famous H-theorem. In doing so, he used the
fundamental classical equation of motion and rigorous mathematical
method. At that time, his work was almost perfect, except that it
was based on a model of colliding particle system for the
macroscopic matter, which is not general enough even from the view
point of classical physics. Now, more than a century passed, some
improvement is needed. We should abandon any special model for the
matter, in order to make the proof general. The quantum theory must
be fully considered. It means we should put the proof entirely on
the basis of fundamental principle only, which should be true both
in classical and in quantum physics. This is what we have done in
our proof.

This work is supported by the National Nature Science Foundation of
China with Grant number 10875003.

\b{\t}{99} \bi{1}  Q.-R. Zhang  A general information theoretical
proof for the second law of thermodynamics Int. J. Mod. Phys. E{\bf
17} 531 (2008)\bi{2} Q.-R. Zhang Axiomatic foundations for the
principle of entropy increase Int. J. Mod. Phys. E{\bf 17} 1075
(2008) \bi{3}A. Feinstein {\it Foundation of Information Theory}
(New York, London; McGraw-Hill 1958)\bi{l} L. D. Landau and E. M.
Lifshitz {\it Statistical Physics}, 3rd edition (Butterworth
Heinemann, Oxford, 1980) \bi{4}T. Stonier Information as a basic
property of the universe Biosyst. {\bf 38} 135 (1996) \bi{5} G. M.
Wang, E. M. Sevick, E. Mittag, D. J. Searles, and D.J. Evans
Experimental demonstration of violations of the second law of
thermodynamics for small systems and short time scales Phys. Rev.
Lett. {\bf 89} 050601 (2002) \bi{6}D. J. Evans, E. G. D. Cohen, G.
P. Morriss  Probability of second law violations in shearing steady
states Phys. Rev. Lett.{\bf 71} 2401 (1993) \e{\t}
\end{document}